\documentclass{article}

\usepackage{arxiv}

\usepackage[utf8]{inputenc} 
\usepackage[T1]{fontenc}    
\usepackage{hyperref}       
\usepackage{url}            
\usepackage{booktabs}       
\usepackage{amsfonts}       
\usepackage{nicefrac}       
\usepackage{microtype}      
\usepackage{lipsum}

\usepackage{float}
\usepackage{amsmath} 
\usepackage{nccmath}
\usepackage{amssymb}  
\usepackage{graphicx}
\usepackage[colorinlistoftodos]{todonotes}
\usepackage{authblk}

\usepackage{longtable}
\usepackage{array}
\usepackage{tabularx}
\usepackage{tabu}
\usepackage{multirow}
\usepackage[flushleft]{threeparttable}

\newenvironment{conditions}
  {\par\vspace{\abovedisplayskip}\noindent\begin{tabular}{>{$}l<{$} @{${}={}$} l}}
  {\end{tabular}\par\vspace{\belowdisplayskip}}

\title{Forecasting of the Montreal Subway Smart Card Entry Logs with Event Data}

\author[1,2]{Florian Toqué}
\author[1]{Etienne Côme}
\author[2,3]{Martin Trépanier}
\author[1]{Latifa Oukhellou}

\affil[1]{COSYS-GRETTIA, Université Gustave Eiffel, IFSTTAR, F-77447 Marne-la-Vallée, France}
\affil[2]{Polytechnique Montréal, Department of Mathematics and Industrial Engineering, Montréal, Canada}
\affil[3]{Interuniversitary Research Centre on Entreprise Networks, Logistics and Transport (CIRRELT), Montr\'eal, Canada}

\begin{document}
\maketitle

\begin{abstract}
One of the major goals of transport operators is to adapt the transport supply scheduling to the passenger demand for existing transport networks during each specific period. Another problem mentioned by operators is accurately estimating the demand for disposable ticket or pass to adapt ticket availability to passenger demand. In this context, we propose generic data shaping, allowing the use of well-known regression models (basic, statistical and machine learning models) for the long-term forecasting of passenger demand with fine-grained temporal resolution. Specifically, this paper investigates the forecasting until one year ahead of the number of passengers entering each station of a transport network with a quarter-hour aggregation by taking planned events into account (e.g., concerts, shows, and so forth). To compare the models and the quality of the prediction, we use a real smart card and event data set from the city of Montréal, Canada, that span a three-year period with two years for training and one year for testing.
\end{abstract}


\section{Introduction}
Public authorities currently play a significant role in encouraging sustainable development policies, giving impetus to sustainable urban mobility practices that aim to reduce the use of private cars and increase the use of sustainable transport modes, such as public transport. In urban and peri-urban areas, this transport strategy faces a number of challenges, including the regularity, quality of service and congestion of public transport. One of the major goals of stakeholders (operators and authorities) is to adapt as accurately as possible the schedules to the passenger demand during each specific period (e.g., normal period, period under events, disturbed period, special day, and so on). According to transport operators, another goal is to anticipate the demand for disposable ticket or pass (non-rechargeable smart cards) to match ticket availability to passenger demand during a specific period, particularly event periods (e.g., concerts, sports games, shows, exhibitions, and so forth). Furthermore, this information on the number of type of ticket or pass used per quarter hour can be used to provide mobility services adapted to the different types of passengers (regular/occasional). For example, a larger number of agents may be made available in the event of a high number of occasional passengers to help manage the extra passenger flow.

To address these issues, we propose a generic data shaping of contextual data, allowing the use of well-known regression models for long-term forecasting of passenger demand with fine-grained temporal resolution. In this study, we forecast the number of passengers entering each of the 68 metro stations in the city of Montréal, Canada, until one year ahead by taking calendar information and planned events into account. We also predict passenger demand per type of ticket or pass used to travel to address the problem of adapting ticket availability to passenger demand. The aggregation time window for the number of passengers has been chosen as 15 minutes, which permits the precise analysis of the impact of events on passenger demand and is relevant to adapting transport supply. We compare several well-known forecasting models, including  basic, statistical and machine learning models. In this context, we analyse the use of contextual data such as information about the day and an event database provided by the public transportation authority of Montréal (Société de transport de Montréal, STM). This methodology aims to be reproducible to forecast the passenger demand for other transport networks around the world (depending on the availability of equivalent data sets in the other cities).

The main objective of this study is to determine whether it is possible to predict the number of passengers using the calendar and event information available in advance (in this case, available one year in advance), with the following innovative aspects: 

\begin{itemize}
\itemsep=0pt
\item Predict the number of incoming passengers at each station of a transportation network (68 stations in the Montréal metro network in Canada)
\item Propose a generic data shaping of contextual data
\item Carry out the study over a long period of time (2 years for the learning set and 1 year for the test set)
\item Predict the number of passengers aggregated with a fine temporal resolution (15 minutes)
\item Perform a detailed analysis of the forecasting results during different periods (e.g. event periods and periods without events)
\item Forecast the number of passengers and perform an analysis of this forecast based on the type of ticket or pass used to travel.
\item Compare several forecasting methods, including basic methods, statistical methods and machine learning methods
\end{itemize}
First, early forecasting of the number of aggregated passengers per quarter hour at each station is useful to transport operators to help them improve the planning of the transport supply schedule (e.g., the number of subways per quarter hour, when planning to increase the supply of related transport systems such as buses) to match it as closely as possible to passenger demand. In addition, this demand forecast can be used to plan the presence of agents, secure stations in the case of excessive passenger traffic and allow passengers to avoid overcrowded situations.

 Note that this approach can only be applied on fixed networks. To be effective, the approach requires a historical data set that includes the occurrence of events at a station; otherwise, the forecasting model will not be able to take into account the event information at a station that never hosted an event in the historical database.

The remainder of this paper is organised as follows. Section~\ref{sec:related_work} details the related work. The case study is presented in Section~\ref{sec:casestudy}. Section~\ref{sec:forecasting} details the forecasting methods and the data shaping that we have developed.  Section~\ref{subsec:forecast_results_allpass} describes the forecasting results on the global aggregation of type of ticket or pass used to travel, while Section~\ref{subsec:forecast_results_perpass} provides an analysis of the forecasting performance per type of ticket or pass used to travel. Finally, some possibilities for future research and conclusions are outlined in Section \ref{sec:conclusion}.

\section{Related Work}
\label{sec:related_work}

Since 2004, the use of smart card data to analyse mobility in public transportation has received substantial attention from researchers. More recently, studies on mobility analysis have revolved around passenger demand forecasting. A distinction can be made between research that relates to forecasting OD matrices and research that attempts to forecast passenger flows at a specific point. Knowledge about these two factors is indeed essential for planning, operation and management in any transportation network, but each of these areas uses different types of data.

The passenger demand goals differ depending on the forecasting time horizon. For long-term forecasting, the aim is to forecast demand with data available at the long-term period in advance (e.g., time features and planned events), which can be very useful for improving transport supply scheduling. In contrast, the forecasting process can also account for the last observations, in which case it is generally referred to as short-term forecasting.

Going forward, in the case of an atypical situation, the main goal for transport operators is to use the forecasted passenger demand to optimise transport system operation to match transport supply to the atypical demand or propose to the passenger an alternative way to reach their destination.

\subsection{Short-term Forecasting of Passenger Demand in Public Transport}
Short-term forecasting, which corresponds to a few time steps ahead forecasting, has been studied with different models. \cite{Li2017} used multiscale RBF networks to forecast the number of alighting passengers at different Beijing subway stations multiple time steps ahead (t+15 and t+30 minutes) by taking the number of boarding passengers at the other station of the subway network into account. In this study, the authors performed an in-depth analysis of the results obtained under special event scenarios. Other examples of subway passenger flow forecasting include the work of \cite{Roos2016}, where the authors predicted passenger flows of the next time step (t+2 minutes). The authors used a Bayesian network model and predicted multiple passenger flows (entry and exit) at all the stations of a subway line of the Paris network. In the study of~\cite{Cui2016}, the authors created a fuzzy nonlinear autoregressive exogenous model to predict the number of passengers at the next time step (t+1 hour). In addition to forecasting, \cite{Ding2016} conducted an in-depth analysis of the influence of subway predictor variables, such as bus transfer activities, and temporal features on the forecasting results and showed that the most important short-term forecasting features are the past observation of the metro ($\sim82.0$\%), the past observation of the bus ($\sim10.4$\%) and the prediction time step ($\sim3.6$\%). This study predicted the next time step (t+15 minutes) of passenger flows at 3 stations of the Beijing subway network.

\subsection{Short-term Mobility Forecasting with Spatiotemporal Focus}
A closer examination of the most recent studies about short-term forecasting in the transportation field reveals  high spatial and temporal values in such prediction problems. For example, in ride-sharing demand forecasting, a research team from Uber (\cite{Laptev2017}) studied Uber ride-sharing demand data with a focus on the temporal values for extreme event forecasting. \cite{ke2017} focused  on capturing knowledge from the spatiotemporal information of the ride network via a deep learning approach. Similar approaches have been performed by~\cite{Zhang2017} to predict citywide crowd flows and by~\cite{Yao2018} to predict taxi demand. Studies that spotlight the spatiotemporal aspect of traffic forecasting have also been conducted by~\cite{Wu2016,cheng2017} with a combination of convolutional and recurrent neural network models and by~\cite{yu2017} with a graph convolutional neural network model.

\subsection{Event Data Usage in Short-term Forecasting}
Some studies have shown the importance of external data, especially event data, for improving the prediction accuracy of forecasting models. Events such as concerts, shows, and sports games are sources of disturbance regarding human mobility. \cite{Ni2017} developed short-term prediction approaches to forecast subway passenger flows for the next 4 hours using social media data. The authors focused on predicting the total number of passengers (sum of entry and exit) of one subway station of the New York City network. They proposed a two-step methodology: hashtag-based event detection followed by the combined use of linear regression and a seasonal autoregressive moving average model. More recent studies conducted by \cite{Markou2018, Rodrigues2019}  involved automatic event data collection, where the authors worked on the short-term forecasting of taxi demand in two distinct locations in New York city by using deep learning methods. In these studies, the model comparison showed that event categorisation could significantly help forecasting models obtain better results.

As shown in Table~\ref{tab:relatedwork_shorttermforecasting}, numerous studies consider short-term forecasting with various methods and forecasting horizons.

\begin{table}[!htb]
\centering
\caption{Related work on short-term forecasting}\label{tab:relatedwork_shorttermforecasting}

\begin{threeparttable}

\begin{tabularx}{1.\textwidth}{p{38mm}Xp{18mm}p{24mm}XX}
\hline
Reference & Method & Mode & Aggregation&Horizon & Event  \\
\hline
\cite{Li2017} &RBF&Subway&15 min&1,2& No\\
\cite{Roos2016} &Bayesian&Subway&2 min&1&No\\
\cite{Cui2016} &AR&Subway&1 h&1&No\\
\cite{Ding2016} &MLP&Subway&15 min&1&No\\
\cite{Laptev2017} &LSTM&Taxi&1 day&1&No\\
\cite{ke2017} &CRNN&Taxi&1 h&1&No\\
\cite{Zhang2017} &CRNN&Taxi\&Bike&1~h \& 30~min&1\&1&No\\
\cite{Yao2018} &CRNN&Taxi&30 min&1&No\\
\cite{Wu2016} &CRNN&Traffic&5 min&1&No\\
\cite{cheng2017} &CRNN&Traffic&15 min&1,2,3,4&No\\
\cite{yu2017} &GCNN&Traffic&15 min&1,2,3&No\\
\cite{Markou2018} &GP &Taxi&1 h&1&Yes\\
\cite{Rodrigues2019} &LSTM&Taxi&1 day&1&Yes\\
\hline
\end{tabularx}
\begin{tablenotes}
    \scriptsize
    \item RBF represents radial basis function network. AR represents autoregressive method. MLP represents multilayer perceptron. LSTM represents long short-term memory introduced by \cite{hochreiter1997}. CRNN represents a different architecture of neural network with convolution and recurrent neural network. GP represents Gaussian process. GCNN represents graph convolutional network.
  \end{tablenotes}
\end{threeparttable}
\end{table}

\subsection{Long-term Passenger Demand Forecasting}
To the best of our knowledge, only a few resources related to long-term forecasting with fine-grained resolution are available in the literature, unlike short-term forecasting. The study most related to our work is the study of \cite{Pereira2015}. The authors worked on long-term forecasting approaches using event data extracted from the web as features to forecast the aggregated number of passengers per half hour of tap in/out of 3 subway and 11 bus stops assigned to 5 venues in the city of Singapore. Their study was performed on a data set with a total period of 16 days. They  demonstrated that using event information (online information) combined with public transport data can improve the quality of transport prediction under special events.

In this study, we investigate the problem of long-term (one year ahead) passenger demand forecasting, represented as the number of tap ins aggregated by 15-minute intervals of all the subway stations (68 stations) in the city of Montréal, Canada, and the use of an event database given by the transport organisation of Montréal. The real data set spans a long period (3 years). We propose a data shaping method that allows the use of well-known regression models for long-term forecasting of passenger demand. Moreover, we study the forecasting of the passenger demand per type of transit fare to provide an in-depth analysis of the passenger demand forecasting, thus helping transit operators adapt the availability of specific pricing during special events.


\section{Case Study}
\label{sec:casestudy}
The forecasting of transport demand at each station of a public transport network is a challenging task, mainly due to the influence of several well-known factors introduced by~\cite{Zhang2017} on crowd flows and on transport demand. These factors can be summarised as follows: temporal factors, including time interval and the type of day, i.e., Monday, Tuesday, ..., Sunday; public or school holidays; and extra day off. Spatial factors include the type of area where the station is located (e.g., residential, office, shopping, and areas of interest). Predictable factors include weather, events, transport operator strikes and renovations. Unpredictable factors include transport network disruption that could be induced by a technical problem (rail problem, fire accident), a passenger problem or another factor that could severely impact the transport supply.

In this study, we aim to perform one-year-ahead forecasting by taking the temporal, spatial and contextual factors into account. To this end, temporal and contextual data that are available one year ahead will be used as inputs of the forecasting models. In the following sections, we detail the smart card entry logs, the time features and the event database.

\subsection{Smart Card Entry Logs}
\label{subsec:smartcardentrylogs}
The real dataset used to evaluate the proposed methodology was provided by the transport organisation authority of Montréal, Canada (Société de transport de Montréal, STM). The ticketing logs used in our study are obtained thanks to the validation of passes and tickets on automated fare collection systems for each user's entry into the transport network. We address all 68 subway stations in the city.  The data set consists of ticketing logs aggregated by 15-minute intervals during 2015, 2016 and 2017. The studied subway network handles more than 670k passengers every day. We also forecast the number of passengers by the type of ticket or pass used to travel. We have aggregated the passengers according to their type of ticket or pass: STM monthly pass, regional monthly pass, book tickets and occasional passes. Disposable tickets include tickets used occasionally (occasional passes), 1- or 2-way tickets, 1- or 3-day passes, weekend passes, special event tickets, etc.

\begin{figure}[!htb]
    \centering
    \includegraphics[height=10em]{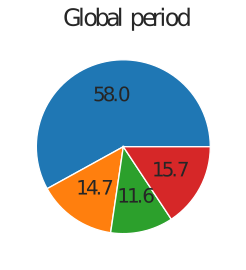}
    \hspace{-1em}
    \includegraphics[height=10em]{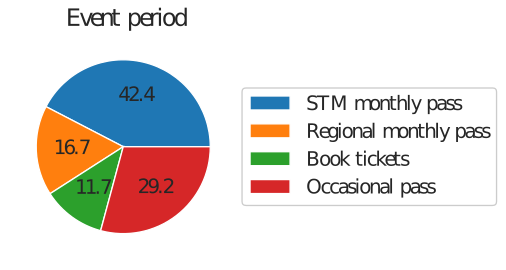}
    \caption{Use of the type of ticket or pass in percentage (2015-2017): global passenger demand on the left, passenger demand during event period on the right (pairs of days and stations with events)}
    \label{pie_chart_pass_use_daystation-event}
\end{figure}

From Figure~\ref{pie_chart_pass_use_daystation-event}, we can see the percentage of passengers entering the subway network according to their type of ticket or pass during the global period from 2015-2017 and during the event period (pairs of days and stations with events). During the global period, the most used pass is the STM monthly pass, with approximately 140M entries per year, which represents 58\% of the passenger demand. On the other hand, during event periods, the percentage of occasional passes increases significantly to 29.2\%, versus 15.7\% during the global period.

\subsection{Detailed Calendar Information}
\label{subsec:dayinformation}
Passenger demand mainly depends on the day type; therefore, we created a list of nine day features, as follows:
\begin{itemize}
    \itemsep=0pt
    \item Name of the day of the week (e.g., Monday, Tuesday, and so on)
    \item Month (e.g., January, February, and so on)
    \item Holiday (e.g., Christmas day, New Year's day and so on.)
    \item 24th of December
    \item 31st of December
    \item Christmas holiday
    \item Summer university holiday part 1 (intensive session, Université de Montréal)
    \item Summer university holiday part 2 (regular session, Université de Montréal)
    \item Renovation period that took place at the  Beaubien station over 4 months in 2015
\end{itemize}  

This list of features is certainly specific for this transport network and this city, but it could easily be modified to suit another transport network and city.

\subsection{Event Data}
\label{subsec:eventdata}

Passenger demand strongly depends on different contextual factors. Some factors cannot be planned far in advance (e.g., weather, transport network disruptions, and so forth), whereas others can be planned in advance, such as the presence of large events in a city (sports games, festivals, concerts, and so on). We could manually create or even automatically extract such event databases, as shown in previous studies conducted by \cite{Moreira2016, Markou2018, Rodrigues2019}. In our case, a real data set of events was provided by the STM operator, who manually built a calendar of events in the city of Montréal occurring during the three years of the study.
 Each event is characterised by a location (the nearest subway station), the event start and end times (format is "Y-m-d H:M:S", approximately 80\% of the event end times are available) and a manually built short text description of the event (description does not follow the same construction pattern, e.g., the same event could have different descriptions). We manually defined 10 topics as event categories to have an event categorisation able to be taken into account by the forecasting models. Taking these type of data into account is a challenge because it involves a large and sparse encoding of the data that is difficult to treat with regression models. Moreover, the end time of the event is not available for each event, which makes the interpretation of the event difficult.

\begin{figure}[!htb]
\centering
\includegraphics[width=0.80\textwidth]{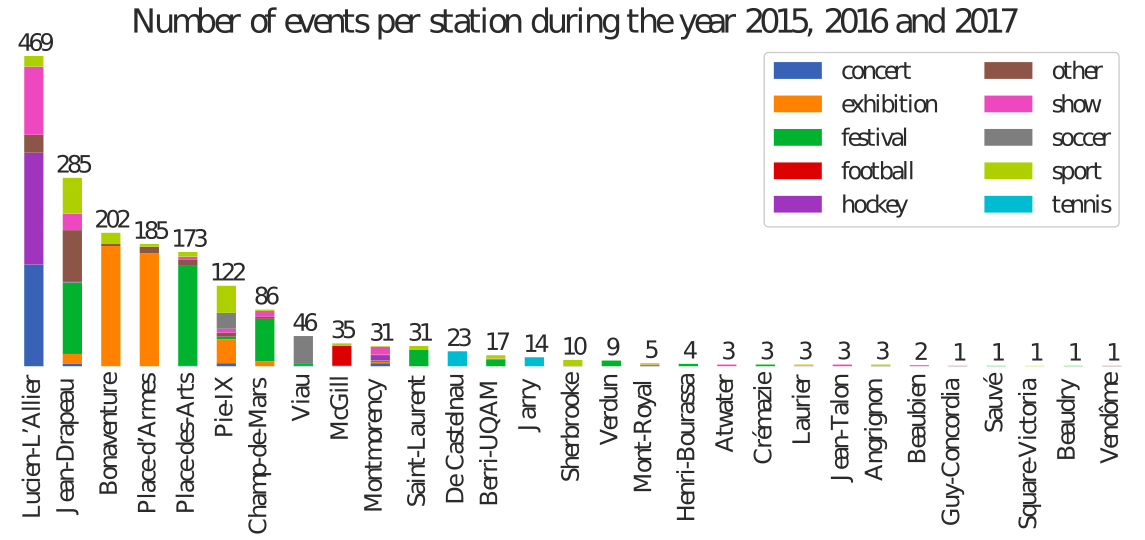}
\caption{Number of events per metro station that host the event and per category.} \label{fig:numberofevents_station}
\end{figure}

Figure~\ref{fig:numberofevents_station} shows the number of events per station and per category. We consider an event by the presence of a start time in the database (e.g., if the same event occurs on 4 consecutive days and is represented by 4 start times in the database, it will be counted as 4 events). As shown, most of the events occur near three stations: Lucien-L'Allier, Jean-Drapeau and Bonaventure.

\begin{figure}[!htb]
\centering
\includegraphics[width=0.80\textwidth]{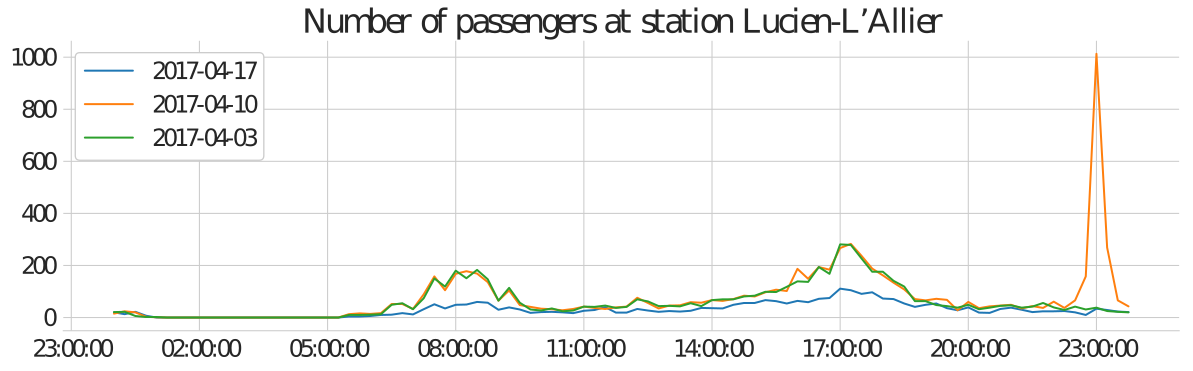}
\caption{Number of passengers on three different Mondays. April 3, 2017, corresponds to a normal Monday; April 10, 2017, is a Monday with an event (Def Leppard concert that finished at 11:00 p.m.); and April 17, 2017, is a holiday (Easter Monday).} \label{fig:observation_3monday}
\end{figure}

To provide an overview of the smart card data set and the differences in passenger demand that could occur between the same type of day with or without the presence of an event, the numbers of passengers on three different Mondays of the same month (April 2017) at the station named "Lucien-L'Allier" are depicted in Figure~\ref{fig:observation_3monday}. Monday, April 3, 2017, is depicted by the green line and could be considered  a normal Monday. We can observe the typical morning and evening peaks of  passenger demand. Monday, April 10, 2017, is coloured in orange, and this day is considered special because an event (Def Leppard concert that finished at 11:00 p.m.) occurred on this day near this station. Finally, Monday, April 17, 2017, which is a holiday (Easter Monday), is depicted by the blue line. We can observe a decrease in passenger demand throughout all Easter Monday (blue) compared to the normal Monday (green). However, we can see a highly concentrated increase in passenger demand due to the end of the concert during the Monday with the event (orange).

\section{Forecasting Workflow}
\label{sec:forecasting}

We aim to forecast the number of passengers entering each station of a transport network at each time step of a day until one year ahead. Here, we forecast the passenger demand of the 68 metro stations in the city of Montréal, Canada, at each quarter hour of a day (96 time steps per day) by taking planned events into account. We have compared the use of different sets of features as inputs of the forecasting models and different types of forecasting models.  Section~\ref{subsec:configuration} details the data shaping and the compared set of features. The general description of the compared models is depicted in Section~\ref{subsec:methodology} and the evaluation metrics are described in Section~\ref{subsec:evaluationmetrics}.

\subsection{Data Configuration}
\label{subsec:configuration}
To evaluate the importance of each contextual data set, we trained the forecasting models with four input data sets (D1, D2, D3 and D4). Each of these input data sets corresponds to a specific concatenation of the following 4 sets of features:
\begin{itemize}
\itemsep=0pt
\item A: Month and name of the day of the week, encoded as one-hot vectors.
\item B: Holiday, 24 of December and 31 of December, Christmas school holiday, university holidays part 1 and part 2, and Beaubien station renovation period. These features are encoded as one-hot vectors.
\item C: Start event, end event and period event at each station that hosts an event. For each station that hosts an event (29 stations), at each time step of the day (vector size 96), we counted the number of time steps related to event schedule information (3 features), namely, the start time, end time and event period. For example, if we encode this information for one station that hosts an event on the day from 00:00 a.m to 00:45 a.m. We will obtain the following three vectors of size 96: (i) start time $[1, 0, ..., 0]$, (ii) end time $[0, 0, 1, 0, ..., 0]$ and (iii) event period $[1, 1, 1, 0, ..., 0]$.
\item D: Category of the event (10 event categories). This has the same encoding as C, but the difference is that we counted the number of event per category. For each station that hosts an event (29 stations), at each time step of the day (vector size 96), for each category of event (10 categories), we counted the number of time steps related to event schedule information (3 features), namely, the start time, end time and event period.
\end{itemize}

The input data sets D1, D2, D3 and D4 are depicted in Table~\ref{tab:setoffeatures}.

\begin{table}[!htb]
\centering
\caption{Input data sets D1, D2, D3 and D4.}\label{tab:setoffeatures}

\begin{threeparttable}

\begin{tabularx}{0.98\textwidth}{cXXXXp{55mm}}
\hline
Data & A & B & C & D & Size\\
{D1}& \checkmark & & & & $11+6=17$ \\
{D2}& \checkmark&\checkmark & & & $17+7=24$\\
{D3}& \checkmark & \checkmark & \checkmark & & $24+96\times3\times29 = 8376$\\
{D4}& \checkmark &\checkmark & \checkmark & \checkmark & $8376 + 96\times3\times29\times10 = 91896$ \\
\hline

\end{tabularx}
\begin{tablenotes}
    \scriptsize
    \item The set of features: A corresponds to the month and name of the day of the week, B corresponds to the detailed day features, C corresponds to the event features, and D corresponds to the category of the event features.
  \end{tablenotes}
\end{threeparttable}
\end{table}

\begin{figure}[!htb]
\centering
\includegraphics[width=0.65\textwidth]{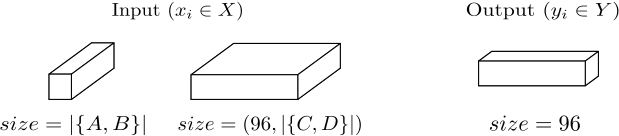}
\caption{Data shaping of one input sample ($x_i \in X$) composed of features \{A,B\} (day features) and \{C,D\} (event features) and of one output sample ($y_i \in Y$) that corresponds to the forecasting of the number of passengers entering a station at each of the 96 time steps. } \label{fig:input_output}
\end{figure}

We have trained a specific forecasting model per station (total of 68 models) with daily multi-time-step output forecasting. This means that for each day, we perform a unique prediction that corresponds to a vector containing the forecasting of the number of passengers per quarter-hour intervals (output vector size is equal to 96, the number of 15-minute time steps in 24 hours). All the forecasting models (one model per station) have the same inputs and outputs, which are depicted in Figure~\ref{fig:input_output}. This figure depicts one input sample ($x_i \in X$) composed of features \{A,B\} and \{C,D\} corresponding to the features available until one year in advance of the forecasted $day_i$. Features A and B are detailed in Section~\ref{subsec:dayinformation} (e.g., day of the week, holiday, school holiday, and so forth). Meanwhile, features C and D are encoded per time step (96 quarters of hour per day) and correspond to the event features.

\subsection{General description of the compared models}
\label{subsec:methodology}

We aim to forecast the passenger demand until one year ahead with a fine-grained temporal resolution (quarter-hour aggregation). In this context, it is not possible to use and optimise the parameters of time series forecasting as autoregressive models (ARIMA, SARIMAX, and so forth) because of the too large training data set and the multi-time-step ahead prediction. Therefore, we compared different well-known models that could be used for regression problems. We computed a baseline model based on a historical average, a linear regression model, machine learning models such as random forest, gradient boosting decision trees, artificial neural network and kernel-based models including a support vector regressor and Gaussian process.

\subsubsection{Historical Average}
The historical average model is a baseline model that aims to predict passenger flows based on the average value of historical observations by type of day in the corresponding time step. The type of day could be defined by a set of features, such as those depicted in Section~\ref{subsec:dayinformation}. For example, one may take the most basic feature, which is the name of the day of the week. In this case, the prediction for the time step of 8:00 a.m.-8:15 a.m., a Monday, corresponds to the average of all the historical values for Monday at 8:00 a.m.-8:15 a.m. The model computes the average number of passengers from the available historical dataset, representing two years in our case. This approach is the baseline of the long-term forecasting models.

\subsubsection{Linear Regression}
Linear regression is a statistical model that assumes a linear relationship between the dependent variable and one or more explanatory variables (or independent variables). We use as input more than one explanatory variable, and we predict more than one dependent variable (total of 96 dependent variables). In this context, we choose a multivariate linear regression. To prevent the collinearity phenomenon with categorical features, such as the name of the day of the week, we formatted the data by deleting one of the categories. To avoid overfitting, we computed the linear regression with the elastic net regularisation introduced by \cite{Zou2005} that linearly combines the L1 and L2 penalties of the lasso and ridge methods.
We optimised the hyperparameters alpha and l1\_ratio, where alpha is a constant that multiplies the penalty terms L1 and L2, and l1\_ratio corresponds to the penalty term associated with the L1 method and (1-l1\_ratio) to the L2 method.

\subsubsection{Random Forest (RF)}
\label{subsubsection:randomforest}
Random forest is a well-known machine learning model whose effectiveness for performing regression or classification problems has been widely proven for many real-world applications. The model introduced by \cite{breiman2001} is an ensemble learning algorithm based on the average prediction of different decision trees (forest). Each tree is fit on different parts of the data, which were created by applying two sampling methods: random sampling with replacement, which is also known as the bootstrap aggregation or bagging method, and random selection of features, which is called feature bagging. The bagging methods and the averaging of the results obtained by the different trees make the RF more robust and accurate than a simple decision tree. We optimised the hyperparameters n\_estimators, which corresponds to the number of trees used by the model; min\_samples\_split, which corresponds to the minimum number of splits required to split an internal node; min\_samples\_leaf, which is the minimum number of leaves required to be at a leaf node; and max\_features, which is the number of features to consider when searching for the best split.

\subsubsection{Gradient Boosting Decision Tree (GBDT)} Gradient boosting, introduced by \cite{Friedman2001}, is a machine learning model for regression or classification tasks that uses an ensemble of weak prediction models, such as decision trees in our case, to create a prediction model. Similar to most of the other boosting methods, GBDT builds weak learners (decision trees) one at a time, where each new tree helps to correct the errors made by previously trained trees. After a tree is added, the data weights are readjusted. Correctly classified input data lose weight, and misclassified examples receive a higher weight. This technique helps future trees focus more on input data that were misclassified by previous trees. We optimised the same hyperparameters as the random forest model detailed in Section~\ref{subsubsection:randomforest}.

\subsubsection{Artificial Neural Network (ANN)}
An artificial neural network, also known as a neural network, is a computational model based on the structure and functions of biological neural networks. Each neuron receives inputs and biases, multiplies them by their weights, sums them and combines that sum with their internal state (activation function) to produce an output. In our case, we used the rectified linear unit (relu) function as the activation function of the hidden layer neurons, and the identity function for the neurons of the last layer (used as default by the scikit-learn library for regression problems). We optimised the number of layers and the number of neurons per layer, and we used the early stopping technique in order to stop the training of the model automatically.

\subsubsection{Gaussian Process Regressor (GP)}
\cite{Rasmussen2003}  developed the Gaussian process, which is a generic supervised learning method; more specifically, it is a kernel method designed to solve regression problems. The prediction is probabilistic (Gaussian) and interpolates the observations. One of the advantages of this model is that it is able to compute confidence intervals in addition to the prediction. The main disadvantage of Gaussian processes is that they lose efficiency in high-dimensional space and that they use the entire sample of feature information to perform the prediction, which could lead to overfitting. We optimised the hyperparameter alpha, which specifies the noise level in the targets.

\subsubsection{Support Vector Regressor (SVR)}
The support vector regressor is a supervised machine learning model and is based on the kernel method introduced by \cite{Drucker1997}. This model can efficiently perform a nonlinear regression using  the kernel trick, implicitly transforming the data into a high-dimensional feature space to make it possible to perform the linear regression. The implementation is based on support vector machine, which is effective in high-dimensional spaces. We optimised the hyperparameters kernel; gamma, which corresponds to the kernel coefficient; and C, which is the penalty parameter of the error term.

\subsubsection{Trend Factor}
\label{subsec:Trend factor}

\begin{figure}[!htb]
\includegraphics[width=0.98\textwidth]{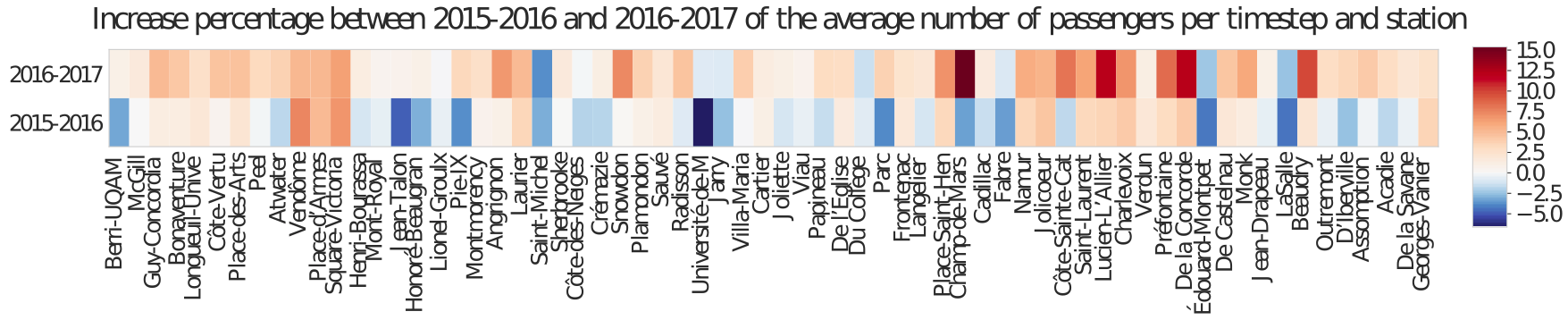}
\caption{Trend factors between 2015-2016 and 2016-2017 per station. Note that 60\% of the stations have the same sign of trend factor between 2015-2016 and 2016-2017.} \label{fig:heatmap_trend}
\end{figure}

The main disadvantage of the forecasting method described in this study is that the models do not take into account the global trend of the number of passengers from year to year.  The heatmap  in Figure~\ref{fig:heatmap_trend} shows the percentage increase between the years 2015 and 2016 and between 2016 and 2017 of the average number of passengers per time step and per station (we do not take into account the Beaubien and Rosemont stations, which were severely impacted by renovations in 2016). As shown, for 60\% of the stations, the increase is of the same sign (positive or negative) between 2015-2016 and 2016-2017.

To take this trend into account in the forecast, we multiplied the forecasted passenger demand at each time step by the trend factor depicted in Equation~\ref{eq:trend_factor}, obtained between 2015 and 2016 (training set). We set the trend factor of the Beaubien and Rosemont stations to 1 to not take the trend factor of these stations into account. To adjust the forecast of the number of passengers per type of ticket or pass used to travel, we calculated a specific trend factor between 2015 and 2016 for each type of ticket or pass used to travel.
\begin{ceqn}
\begin{align}\label{eq:trend_factor}
    trend\_factor_{2015-2016}(s) = \frac{\frac{1}{T_{2016}} * \sum_{t_1=0}^{T_{2016}} x^{t_1}_{2016}(s)}{\frac{1}{T_{2015}} * \sum_{t_2=0}^{T_{2015}} x^{t_2}_{2015}(s)}
\end{align}
\end{ceqn}
where
\begin{conditions}
x & Number of passengers\\
T_y & Number of time step of year $y$, with $t \in T$\\
s & Station $s$
\end{conditions}

 This first attempt to introduce a trend factor in the forecasting model is basic. Further investigations are needed to improve the forecasting capability of the models.

\subsection{Evaluation Methods}
\label{subsec:evaluationmetrics}
To evaluate the models, we split the entire dataset into two different parts: (i) a training dataset used to fit the models with data spanning the years 2015 and 2016 and (ii) a testing dataset used to compare the performance of the models with data of the year 2017. We evaluate the results obtained by the different forecasting models with several well-known metrics. To obtain a better understanding of the errors, three measures of prediction accuracy were used, namely, the root mean square error (RMSE), the median absolute error (MAE) and the mean average percentage error at $v$ (MAPE@v).

These errors can be expressed as follows:

\begin{ceqn}
\begin{equation}
	\operatorname{RMSE}=\sqrt{\frac{\sum_{s=1}^S\sum_{t=1}^T (\hat{y}_s(t) - y_s(t))^2}{T\times S}}
\end{equation}
\end{ceqn}

\begin{ceqn}
\begin{equation}
	\operatorname{MAE}={\frac  {1}{T\times S}}\sum_{s=1}^S\sum_{{t=1}}^{T}\left|\hat{y}_s(t) - y_s(t)\right|
\end{equation}
\end{ceqn}

\begin{ceqn}
\begin{equation}
	\operatorname{\forall y_s(t) > v,~MAPE@v}={\frac {100}{T\times S}} \times \sum _{s=1}^{S}\sum _{t=1}^{T}\left|{\frac {y_s(t)-\hat{y}_s(t)}{y_s(t)}}\right|
\end{equation}
\end{ceqn}

 where $\hat{y}_s(t)$ is the forecast value of station $s$ at time step $t$, $y_s(t)$ is the actual value, and $S$ is the station number.

\subsection{Implementation and Optimisation of the Models}
In this section, we detail the setups of the different forecasting models. We discuss the optimisation of the hyperparameters and the library and resources used to build the models.

\subsubsection{Implementation} We used Scikit-Learn developed by \cite{scikit-learn}, a famous Python library, to compute the following models: elastic net regression, Gaussian process regressor, random forest, gradient boosting decision tree, artificial neural network and support vector regressor. We used the MultiOuputRegressor class of Scikit-Learn to perform multi-output forecasting with SVR and GBDT models that are single-output regressors.

\subsubsection{Optimisation} We performed a grid search with 5 random fold cross-validation to optimise all the statistical and machine learning models. We fixed the computation time for the optimisation of each model to a maximum of 2 days. We used the Scikit-learn default hyperparameters for the model GBDT with input data sets D3 and D4 because of the computation time being too long. The tested hyperparameters are presented in Table~\ref{tab:hyperparameters}. The experiments were conducted in parallel on 20 cores.

\begin{table}[!htb]
\centering
\caption{Grid search hyperparameters of the forecasting models.}\label{tab:hyperparameters}
\begin{threeparttable}

\begin{tabularx}{0.98\textwidth}{Xll}
\hline
Model & Hyperparameter& Tested values \\
\hline
\multirow{3}{*} {LR}  & alpha &0.1, 1, 10  \\
                      & l1\_ratio & 0.25, 0.5, 0.75, 1  \\
                      & normalise & True, False   \\
\hline
\multirow{2}{*} {GP}  & alpha &0.1, 0.5, 1 \\
                      & normalise\_y & False, True \\
\hline
\multirow{4}{*} {RF}  & n\_estimators &100, 150, 200 \\
                      & min\_samples\_split & 2, 5, 10  \\
                      & min\_samples\_leaf & 1, 5, 10   \\   
                      & max\_features & 'auto'   \\  
\hline
\multirow{4}{*} {GBDT}& n\_estimators &100, 150, 200 \\
                      & min\_samples\_split & 2, 5, 10  \\
                      & min\_samples\_leaf & 1, 5, 10   \\  
                      & max\_features & 'auto'  \\  
\hline
\multirow{3}{*} {SVR} & kernel &'rbf', 'linear'  \\
                      & gamma & 1, 0.1, 0.01, 0.001 \\
                      & C & 0.001, 0.01, 0.1, 1.0, 10   \\
\hline
\multirow{3}{*} {ANN} & solver & 'adam'  \\
                      & batch\_size & 16 \\
                      & max\_iteration & 5000  \\
                      & early\_stopping & True  \\
                      & hidden\_layer\_sizes & (10), (100), (300), (10,~10), (100,~10), (100,~100), (300,~100)  \\
\hline
\end{tabularx}

\begin{tablenotes}
    \scriptsize
    \item The value 'auto' of the hyperparameter max\_features of the RF and GBDT models corresponds to the total number of features. The kernel 'rbf' of the SVR model corresponds to the radial basis function kernel. These two values corresponds to the default values in the scikit-learn library. Concerning the hyperparameter hidden\_layer\_sizes of the ANN model, the ith element represents the number of neurons in the ith hidden layer.
  \end{tablenotes}

\end{threeparttable}
\end{table}


\section{Forecasting Results and Discussion}
\label{sec:forecast_results}

First, the results of the passenger demand forecasting with an overall aggregation are presented in Section~\ref{subsec:forecast_results_allpass}.
We compare the forecasting results and present some forecast and observation examples during two different periods: a normal period and an event period. Regarding the methodological context, we show which model performs the best in terms of forecast accuracy and the importance of each feature in the forecasting. In a more focused system transport context, we show the forecasting results per station. Then, we detail the results obtained for each type of ticket or pass used to travel in Section~\ref{subsec:forecast_results_perpass}. 

All the results correspond to the forecast obtained by the different models in addition to the trend factor method explained in Section~\ref{subsec:Trend factor}. The trend factor method improves the results by approximately 0.80\%.

\subsection{Forecasting Result Aggregation of All Types of Ticket or Pass}
\label{subsec:forecast_results_allpass}

\subsubsection{Global Forecast Analysis}
\label{sec:globalanalysis}
To obtain a global comprehension of the results obtained by the models using different sets of features as model inputs, we studied the aggregated errors mentioned in Section~\ref{subsec:evaluationmetrics} of all the stations during the entire training and testing. Table~\ref{tab:results_allpass_globalperiod} depicts the RMSE, MAE and MAPE@150 errors on the training and test sets obtained by the forecasting models described in Section~\ref{sec:forecasting}. The models are used to forecast the number of ticketing logs aggregated per 15 minutes. Each model has been computed with different input data sets (D1, D2, D3 and D4) detailed in Section~\ref{subsec:configuration}. We can observe that the best results are obtained with models using the combination of all the input data sets (D4), except for the ANN model, which is not able to capture the event information because of the excessively low number of training samples due to the special formatting of the data, and the Gaussian process, which overfits because of the excessively large and sparse information caused by the additional event and category of the event features. The best prediction model is obtained with the RF model, with 38.53 and 13.13\% for the RMSE and MAPE@150 error, respectively.
The historical average (HA) model is a basic method in terms of its implementation (it calculates the average number of passengers based on the day type). Unlike the SVR and LR models, where the number of parameters corresponds to the number of features, the number of parameters of the HA model corresponds to all possible combinations of features. For example, for the data set D1, the HA model contains 84 parameters (7 days * 12 months). This explains why this model succeeds in obtaining better results than the LR, and SVR models. On the other hand, it becomes more difficult to predict with this model when the number of features increases (data set D2) and even impossible to predict if the number of features is too large (data sets D3 and D4).

We have seen that the random forest model obtains the best forecast results using the D4 data set over the global period. We will see in Table~\ref{tab:error_allpass_eventperiod} that despite the difference in the number of features between D3 and D4, it is preferable to predict the number of passengers with the D4 data set, since the difference in performance may increase depending on the forecast period.

\begin{table}[!htb]

\caption{Errors on the training and test sets of the different forecasting models with different input data sets (D1, D2, D3 and D4).}\label{tab:results_allpass_globalperiod}

\begin{threeparttable}
\begin{tabularx}{0.98\textwidth}{XXXXXXXX}
\hline
 &&\multicolumn{3}{c}{Train (2015-2016)} & \multicolumn{3}{c}{Test (2017)}\\
\hline
 { Data} & { Model} & { RMSE} & {  MAE} & {  MAPE } & { RMSE} &{  MAE} & {  MAPE}  \\
\hline
\multirow{7}{*}
{D1}&HA & 45.41 & 18.07 & 12.69 & 50.36 & 21.47 & 15.28 \\
&LR& 49.71 & 20.51 & 13.87 & 52.04 & 22.46 & 15.51\\
&RF& 46.97 & 18.76 & 13.01 & 50.39 & 21.32 & 15.17\\
&\textbf{GP}& 46.09 & 18.62 & 12.85 & \textbf{50.32} & 21.56 & 15.17\\
&SVR& 55.98 & 23.72 & 14.12 & 57.54 & 25.40 & 15.58\\
&GBDT& 48.21 & 19.44 & 13.35 & 51.19 & 21.77 & 15.82\\
&ANN&49.93 & 20.57 & 14.51 & 53.13 & 22.76 & 16.18\\

\hline
\multirow{7}{*}
{D2}&HA & 32.24 & 13.04 & 9.73 & 44.31 & 19.16 & 13.84\\
&LR& 41.15 & 17.99 & 12.44 & 44.96 & 20.13 & 13.99\\
&\textbf{RF}& 35.12 & 14.66 & 10.65 & \textbf{41.35} & 18.19 & 13.20\\
&GP& 33.68 & 14.04 & 10.30 & 41.42 & 18.54 & 13.42\\
&SVR& 45.67 & 20.42 & 12.59 & 49.15 & 22.47 & 14.04\\
&GBDT& 37.62 & 16.18 & 11.56 & 42.33 & 18.84 & 13.91\\
&ANN& 40.2 & 16.6 & 12.08 & 43.83 & 18.75 & 13.63\\

\hline
\multirow{6}{*}
{D3} &LR& 34.56 & 16.59 & 11.57 & 43.74 & 20.37 & 14.21\\
&\textbf{RF}& 26.79 & 12.67 & 9.29 & \textbf{39.66} & 17.99 & 13.16\\
&GP& 17.13 & 7.00 & 4.91 & 79.71 & 36.39 & 22.66\\
&SVR& 36.83 & 18.93 & 12.32 & 51.11 & 24.98 & 16.18\\
&GBDT& 26.38 & 13.39 & 9.82 & 42.75 & 18.90 & 14.04\\
&ANN& 40.01 & 18.52 & 13.62 & 55.2 & 25.43 & 18.61\\

\hline
\multirow{6}{*}
{\textbf{D4}}&LR& 33.79 & 16.62 & 11.65 & 42.62 & 20.27 & 14.18\\
&\textbf{RF}& 26.60 & 12.63 & 9.29 & \textbf{38.53} & 17.88 & 13.13\\
&GP& 16.90 & 6.96 & 4.85 & 80.71 & 36.98 & 23.06\\
&SVR& 37.04 & 19.20 & 12.36 & 51.14 & 25.35 & 16.37\\
&GBDT& 26.10 & 13.33 & 9.79 & 40.79 & 18.77 & 14.01\\
&ANN& 41.44 & 19.80 & 14.52 & 63.57 & 29.62 & 21.55\\

\hline
\end{tabularx}

\begin{tablenotes}
    \scriptsize
    \item The data are represented by different sets of features (D1, D2, D3 and D4) described in Section~\ref{subsec:configuration}. The different models are described in Section~\ref{sec:forecasting}. The evaluation metrics RMSE, MAE and MAPE@150 are defined in Section~\ref{subsec:evaluationmetrics}.
  \end{tablenotes}
\end{threeparttable}
\end{table}

\begin{figure}[!htb]
\centering
\includegraphics[width=0.75\textwidth]{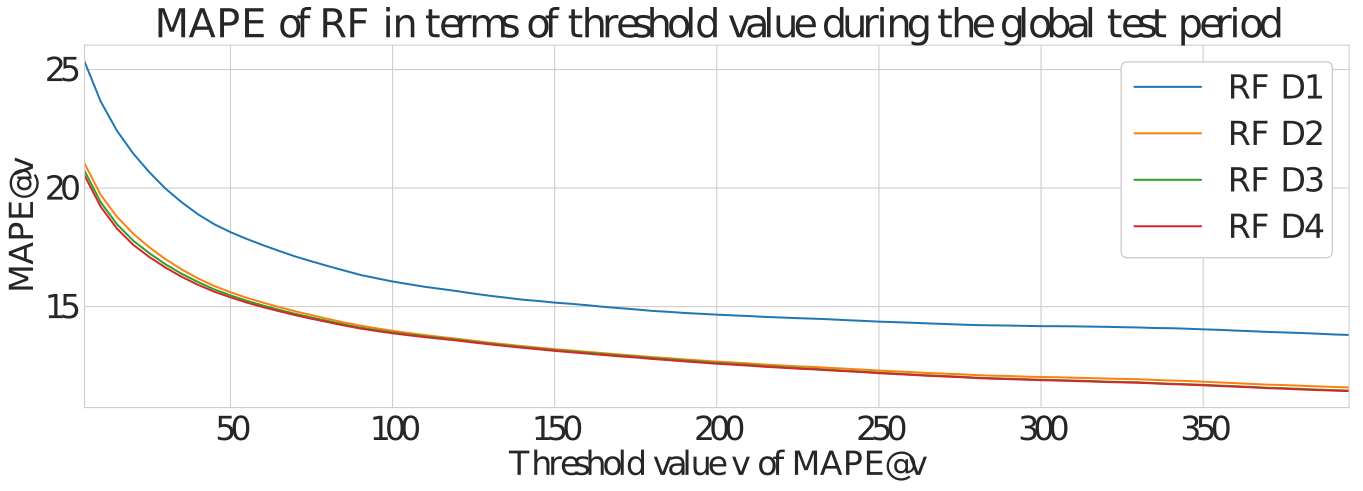}
\caption{MAPE of random forest models with input data sets (D1, D2, D3, D4) in terms of the MAPE threshold value.} 
\label{fig:mapeatv_globaltestperiod}
\end{figure}

As shown in Figure~\ref{fig:mapeatv_globaltestperiod}, the MAPE@v error highly depends on the threshold value. For example, with the best input data (D4), the RF model has a MAPE@5 (MAPE considering all of the observation passenger numbers greater than 5) of approximately 20\% and a MAPE@150 of 13\%. We choose the threshold value of MAPE as 150 to obtain a better estimation of the performance of passenger number forecasting when there is a considerable amount of demand that could impact ticketing demand and transport supply.

\begin{figure}[!htb]
\centering
\includegraphics[width=0.75\textwidth]{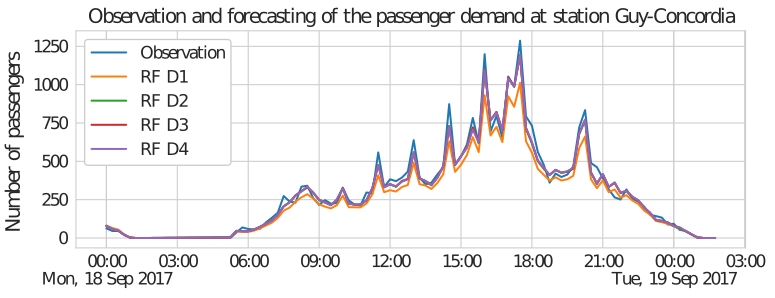}
\caption{Observation and forecasting of the passenger demand at Guy-Concordia station, Monday,  September 18, 2017.} \label{fig:obsvspred_allpass_guyconcordia}
\end{figure}

The observation and forecasting of the random forest model with all the sets of input data (D1, D2, D3 and D4) at Guy-Concordia station are depicted in Figure~\ref{fig:obsvspred_allpass_guyconcordia}. The passenger demand for this station is largely related to the activity of students at Guy-Concordia University. Indeed, we can see that on Monday,  September 18, 2017, the passenger demand appears to follow a regular pattern with activity peaks corresponding to the end of the courses at the university. In this case, the model with input data set D4 succeeds in accurately predicting passenger demand and is slightly better than the models with other input data.

Because events could impact the forecasting results, we analysed the forecasting results of the best forecasting model (RF model) considering periods with and without events (see Table~\ref{tab:error_allpass_eventperiod}). Seventeen stations with events in 2017 (test set period) were extracted. The filtering of the event period is performed by selecting the day/station pair with events. The period without an event represents the remaining data in the considered period. We can observe that the choice of input data significantly impacts the RMSE error during the event period. Indeed, RMSE is slightly improved during the period without an event depending on the use of the input data set D2 or D4 (50.71 against 48.83 of RMSE), whereas this error is largely improved during the event period when D4 is used (153.34 against 124.72 of RMSE). The model with input D1 is too basic to be relevantly compared during periods without events with the model with input D4.

\begin{table}[!htb]
\centering
\caption{Errors of the random forest model applied to the test set period, 2017 (event period and the period without event), on the 17 stations that host events in 2017.
}\label{tab:error_allpass_eventperiod}

\begin{threeparttable}
\begin{tabularx}{0.85\textwidth}{XXXXXXXX}
\hline
 &\multicolumn{3}{c}{Period without event} & \multicolumn{3}{c}{Event period}\\
\hline
 {Data} &{ RMSE} &{  MAE} &{  MAPE }&{ RMSE} &{  MAE}  &{  MAPE}  \\
\hline
{D1}& 61.54 & 28.48 & 14.95 & 159.13 & 46.96 & 23.59\\
{D2}& 50.71 & 24.73 & 13.18 & 153.34 & 43.44 & 22.20\\
{D3}& 49.13 & 24.10 & 13.19 & 137.69 & 43.51 & 21.37\\
{\textbf{D4}}& \textbf{48.83 }& 23.98 & 13.12 & \textbf{124.72} & 40.70 & 21.07\\

\hline
\end{tabularx}

\begin{tablenotes}
    \scriptsize
    \item The data are represented by different sets of features (D1, D2, D3 and D4) described in Section~\ref{subsec:configuration}. The different models are described in Section~\ref{sec:forecasting}. The evaluation metrics RMSE, MAE and MAPE@150 are defined in Section~\ref{subsec:evaluationmetrics}.
  \end{tablenotes}
\end{threeparttable}
\end{table}

\begin{figure}[!htb]
\centering
\includegraphics[width=0.75\textwidth]{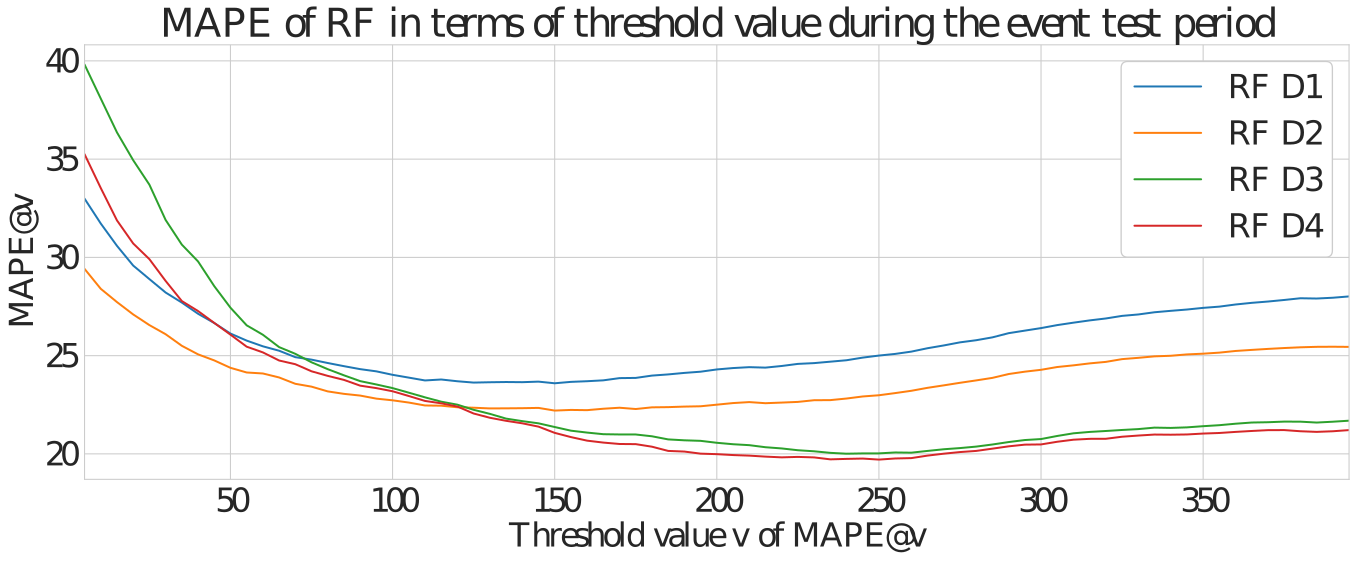}
\caption{MAPE of random forest models with input data sets (D1, D2, D3, D4) in terms of threshold value of MAPE during the event test period.} 
\label{fig:mapeatv_eventtestperiod}
\end{figure}

Figure~\ref{fig:mapeatv_eventtestperiod} depicts the MAPE@v error according to the threshold value v during the event test period of the RF models.
As shown, the best performance is not obtained by the same models with a threshold that is lower or greater than 120, which could be explained by the fact that the calculation of the MAPE@v error is clearly impacted by the passenger number observation. It disadvantages forecasting with a high value when the observation corresponds to a small value over forecasting with a low value when the observation relates to a high value. To improve transport supply and ticket availability in cases of high demand, the usage of a threshold that is greater than a certain value is more relevant than a lower threshold for a model comparison. According to this data set, the threshold of 150 seems to be a good compromise for the evaluation of the models.

\begin{figure}[!htb]
\centering
\includegraphics[width=0.75\textwidth]{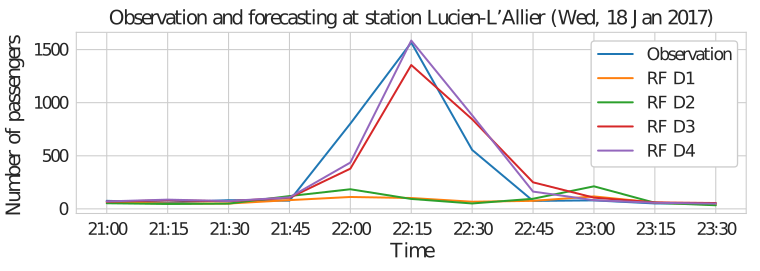}
\caption{Observation and forecasting of the passenger demand at  Lucien L'Allier station, Wednesday,  January 18, 2017. The information about the event is the following: start time, 19:30; end time, 21:30; station, Lucien-L'Allier; and category, hockey.} \label{fig:obsvspred_allpass_lucienlallier}
\end{figure}

Taking the presence of events into account may be essential for forecasting the number of passengers with precision. As shown in Figure~\ref{fig:obsvspred_allpass_lucienlallier}, the random forest with input data set D1 or D2 (detailed information about the day) is not able to predict the high increase in the passenger demand due to the end of a hockey game at  Lucien-L'Allier station. However, with the help of event and event category information (input data set D4), the random forest model accurately forecasts the passenger demand peak.

\begin{figure}[!htb]
\centering
\includegraphics[width=0.75\textwidth]{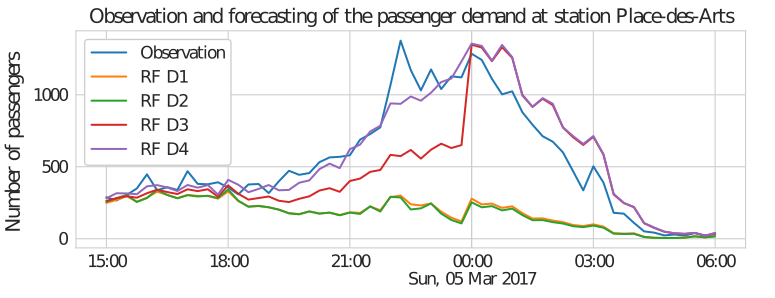}
\caption{Observation and forecasting of the passenger demand at Place-des-Arts station, Sunday,  March 5, 2017. The information about the event is as follows: start date and time, 2017-03-04 18:00:00; end date and time, 2017-03-05 05:00:00;  station, Place-des-Arts; and category, other.} \label{fig:obsvspred_allpass_placedesarts}
\end{figure}

 Figure~\ref{fig:obsvspred_allpass_placedesarts} shows the passenger demand observation during the event named "Nuit Blanche", which induces a very specific pattern due to numerous events occurring during the night in the event area of  Place-des-Arts station and the opening of the metro all night. We can observe a high increase in the number of passengers that has been successfully forecasted by the  random forest model with input data set D4.

\subsubsection{Feature Importance on Specific Station}
\begin{figure}[!htb]
\centering
\includegraphics[width=0.70\textwidth]{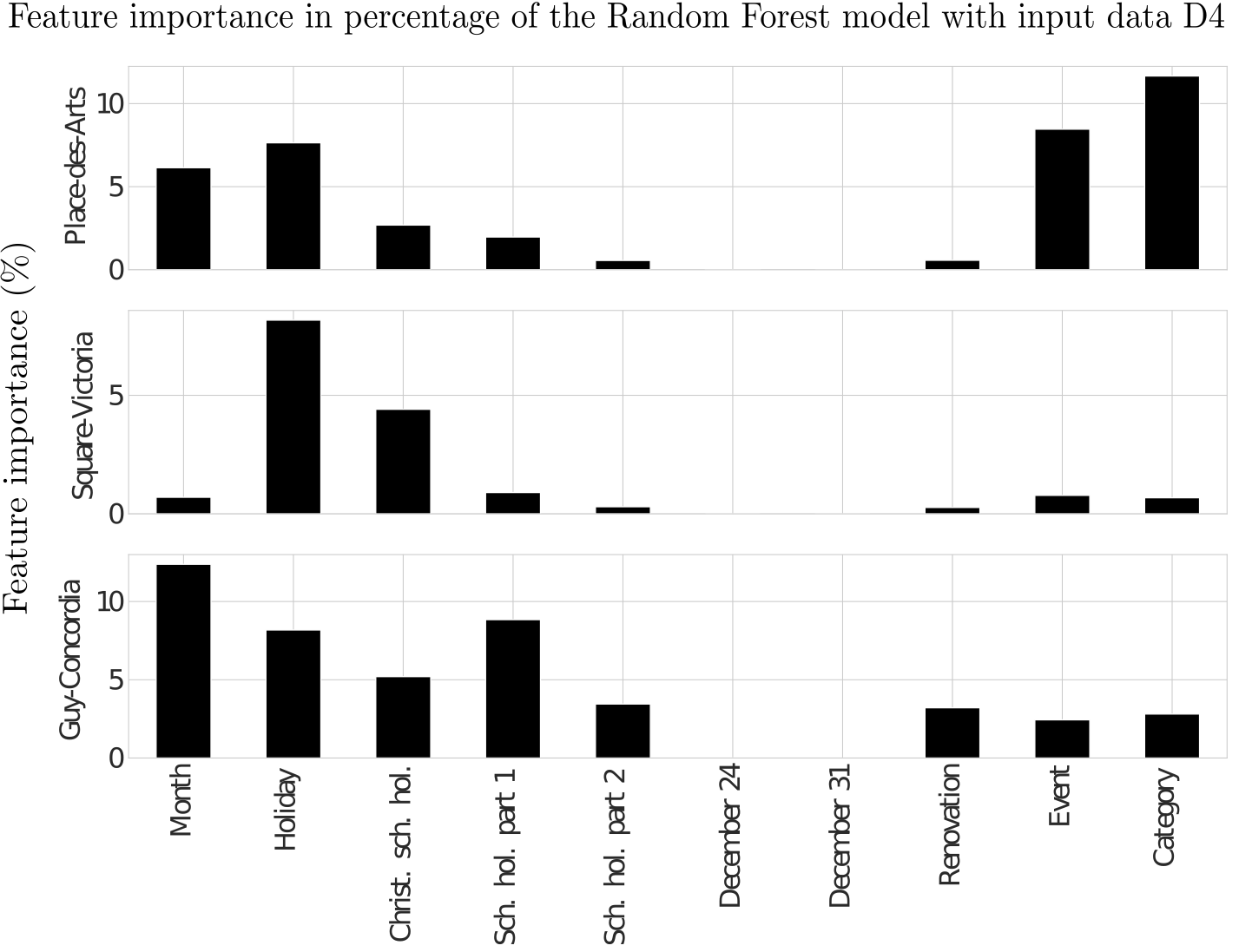}
\caption{Aggregated feature importance of random forest model with input data set D4 in Place-des-Arts, Square-Victoria-OACI and Guy-Concordia stations} \label{fig:feature_importance}
\end{figure}

Models such as random forest allow quantification of the feature importance of the input data. Because one model has been trained per station, we are able to investigate with precision the feature importance per station, which could be interesting for understanding how the models work.  Figure~\ref{fig:feature_importance} shows the feature importance of the random forest model with input data set D4 for 3 stations with particular locations (stations are depicted in Figure~\ref{fig:mape_rf4_test_period}). The feature ranking denoted as $f$ is computed with the "mean decrease impurity" used for regression trees introduced by~\cite{breiman2017}. The importance of feature $i$, denoted as $f_i$, is given by:
\begin{ceqn}
\begin{equation}{\label{equation:gini}}
f_i = \frac{\sum_{j : \text{node j splits on feature i}} n_j}{\sum_{j \in \text{all nodes}} n_j}
\end{equation}
\end{ceqn}
With $n_j$ the importance of node j,
\begin{ceqn}
\begin{equation}
n_j = w_j C_j - w_{left(j)}C_{left(j)}- w_{right(j)}C_{right(j)}
\end{equation}
\end{ceqn}
where $w_j$ is the weighted number of samples in node $j$, $C_j$ is impurity in this node that corresponds to the within node variance of the output value, and left(j) and right(j) are its respective child nodes. The feature importance is given in percentage and has been aggregated in the following categories: the information about the date detailed in Section~\ref{subsec:dayinformation}, events (that corresponds to the sum of the feature importance of all the start, end and period event features of all the stations with events) and category (it corresponds to all the information about the event category available in all the station with events). The most important feature is the name of the day of the week, with 60.31\%, 83.78\% and 53.49\%  feature importance for the Place-des-Arts, Square-Victoria and Guy-Concordia stations, respectively. For these three stations we can see that the importance of the features December 24 and 31 are less than 1\%. This is explained by the few days with these features in the training database (4 days). Nevertheless these features are still important to predict those special days that cannot be categorized with the other features. We can see that Place-des-Arts is a station largely impacted by the event and event category features (approximately 8\% and 11\% of feature importance), which is explained by the presence of many events located near this station. Square-Victoria-OACI is a station located in a business area; in contrast to the Place-des-Arts station, we find that the most important features are holiday and Christmas school holidays. Finally, the Guy-Concordia station is the station of the Guy-Concordia University, which explains the importance of the features Christmas school holidays and school holidays parts 1 and 2.

\subsubsection{Forecast Analysis by Station}
\begin{figure}[!htb]
\centering
\includegraphics[width=0.70\textwidth]{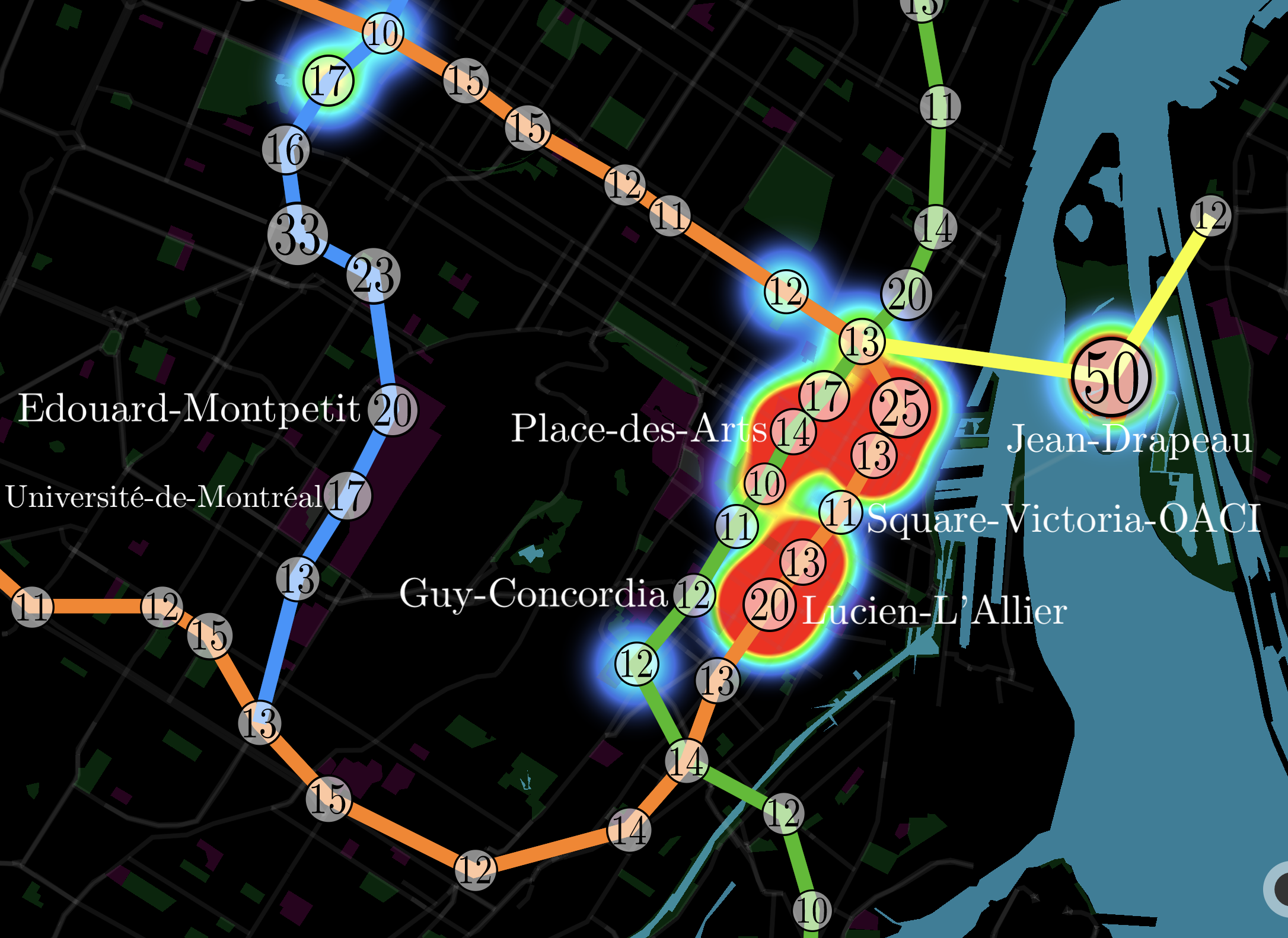}
\caption{MAPE@150 error per station during the global test period (2017) of the random forest model with input D4. The metro lines of Montreal are depicted by the blue, orange, green and yellow lines. The heatmap (blue zone to red zone) depicts the event activity during 2017. The green background represents parks, and the pink background indicates school or university.} \label{fig:mape_rf4_test_period}
\end{figure}

In addition to the global analysis detailed in Section~\ref{sec:globalanalysis}, it is also important to analyse the results per station because each station has its own activity pattern.  Figure~\ref{fig:mape_rf4_test_period} shows the MAPE@150 error of the best forecasting model, which is the random forest with the full set of features as input data (D4 corresponds to the information about the day, the event and the category of the event). As shown, the model obtains an error greater than or equal to 17\% as MAPE@150 in some special stations.
 Université de Montréal and Edouard-Montpetit stations are located on the University of Montreal campus, which implies a passenger demand impacted by the university calendar (MAPE@150 equal to 17\% and 20\%). The Lucien-L'Allier station is difficult to predict (MAPE@150 equal to 20\%) because this station is the one that hosts most of the events in the city. Finally, the several events that took place at the Jean-Drapeau station, located on an island without habitation, make this station the hardest to predict (50\% of MAPE@150).

\subsection{Forecasting Results per Type of Ticket or Pass}
\label{subsec:forecast_results_perpass}

One of the goals of transport operators is to accurately estimate the demand for certain types of ticket or pass used to travel to adapt ticket availability to passenger demand. In this context, we compare the forecasting results of the random forest model with a focus on the forecasting of subsets of data corresponding to the number of passengers by type of ticket or pass used to travel. The type of ticket or pass are aggregated into the following categories: STM monthly pass (SMP), regional monthly pass (RMP), book tickets (BT) and occasional pass (OP).

\subsubsection{Forecast Analysis per Type of Ticket or Pass During the Global Period}
According to the results shown in Table \ref{tab:error_perpass_globalperiod} and as  expected,  the occasional transport demand (pass OP) is the most difficult to forecast. The MAPE@150 is 27.93\% for this type of pass that represents 15.7\% of the total passenger demand against a MAPE@150 of 12.23\% for the type of pass SMP that represents 51\% of the total passenger demand. The use of input data set D4 related to event information in addition to the information about the day is necessary to obtain the best results for the forecasting of occasional passenger demand BT and OP passes. This is due to the particularity of book tickets and the occasional pass that are mainly used during events.

\begin{table}[!htb]
\centering
\caption{Errors of the random forest model on the training period from 2015-2016 and the test set period, 2017, per type of ticket or pass used to travel.
}\label{tab:error_perpass_globalperiod}

\begin{threeparttable}
\begin{tabularx}{0.98\textwidth}{llXXXXXX}
\hline
& &\multicolumn{3}{c}{Train set (2015 and 2016)} & \multicolumn{3}{c}{Test set (2017)}\\
\hline
 { Pass} & { Data} & { RMSE} &{  MAE} &{  MAPE }&{ RMSE} &{  MAE}  &{  MAPE}  \\
\hline
 \multirow{2}{*}{SMP}&\textbf{D2} & 16.37 & 8.65 & 9.67 & \textbf{20.04} & 10.76 & 12.23 \\
& D4 & 14.10 & 7.71 & 8.51 & 20.08 & 10.72 & 12.24\\
\hline
 \multirow{2}{*} {RMP}& \textbf{D2} &8.30 & 3.08 & 9.28 & \textbf{10.17} & 3.76 & 11.97 \\
& D4 & 7.46 & 2.84 & 8.28 & 10.28 & 3.75 & 12.12\\
\hline
 \multirow{2}{*} {BT}& D2& 5.58 & 3.06 & 16.23 & 6.73 & 3.63 & 20.48 \\
& \textbf{D4} & 5.06 & 2.89 & 13.17 & \textbf{6.57} & 3.58 & 19.19\\
\hline
\multirow{2}{*} {OP}& D2 & 19.49 & 4.88 & 28.28 & 21.41 & 5.66 & 30.42 \\
& \textbf{D4} & 11.22 & 4.07 & 18.93 & \textbf{17.86} & 5.40 & 27.93\\
\hline
\end{tabularx}

\begin{tablenotes}
    \scriptsize
    \item The data are represented by different sets of features (D2 and D4) described in Section~\ref{subsec:configuration}. The evaluation metrics RMSE, MAE and MAPE@150 are defined in Section~\ref{subsec:evaluationmetrics}. The aggregation of the types of passes is as follows: STM monthly pass (SMP), regional monthly pass (RMP), book tickets (BT) and occasional pass (OP).
  \end{tablenotes}
\end{threeparttable}
\end{table}

\subsubsection{Forecast Analysis per Type of Ticket or Pass Used to Travel During the Event Period}

The STM monthly pass and regional monthly pass are slightly impacted by events. As shown in Table~\ref{tab:error_perpass_eventperiod}, the RMSE of the 17 stations with events increased from 24.99 to 28.01 during the event period for the STM monthly pass and from 12.48 to 13.27 during the event period for the regional monthly pass. Meanwhile, we can observe that book tickets and occasional passes are highly impacted by the presence of events. The random forest model obtains the best scores for these two types of passes with the input data set D4 in both periods: with and without event periods.

\begin{table}[!htb]
\centering
\caption{Errors of the random forest model on the test event period and test set period without events, 2017, per type of ticket or pass over the 17 stations with events during the year 2017.
}\label{tab:error_perpass_eventperiod}

\begin{threeparttable}
\begin{tabularx}{0.98\textwidth}{llXXXXXX}
\hline
& &\multicolumn{3}{c}{Test period without event} & \multicolumn{3}{c}{Test set period with event}\\
\hline
 {\scriptsize Pass} & {\scriptsize Data} & {\scriptsize RMSE} &{\scriptsize  MAE} &{ \scriptsize MAPE }&{\scriptsize RMSE} &{\scriptsize  MAE}  &{\scriptsize  MAPE}  \\
\hline
 \multirow{2}{*}{SMP}&D2 & \textbf{24.99} & 13.12 & 11.88 & 30.48 & 13.68 & 18.14 \\
& D4 & 25.21 & 13.07 & 11.99 & \textbf{28.01} & 13.27 & 16.49\\
\hline
 \multirow{2}{*} {RMP}& D2 & \textbf{12.48} & 5.27 & 11.20 & 13.47 & 5.67 & 11.18 \\
& D4 & 12.67 & 5.28 & 11.41 & \textbf{13.27} & 5.57 & 11.23\\
\hline
 \multirow{2}{*} {BT}& D2& 8.95 & 4.74 & 17.92 & 16.16 & 6.54 & 51.70 \\
& \textbf{D4} & \textbf{8.80} & 4.65 & 17.44 & \textbf{14.63} & 6.36 & 41.03\\
\hline
\multirow{2}{*} {OP}& D2 & 21.85 & 8.57 & 30.06 & 114.60 & 23.72 & 55.70\\
& \textbf{D4} & \textbf{19.17} & 7.79 & 29.33 &\textbf{ 91.03} & 21.95 & 43.98\\
\hline
\end{tabularx}

\begin{tablenotes}
    \scriptsize
    \item The data are represented by different sets of features (D2 and D4) described in Section~\ref{subsec:configuration}. The evaluation metrics RMSE, MAE and MAPE@150 are defined in Section~\ref{subsec:evaluationmetrics}. The aggregation of the types of passes is as follows: STM monthly pass (SMP), regional monthly pass (RMP), book tickets (BT) and occasional pass (OP).
  \end{tablenotes}
\end{threeparttable}
\end{table}

\begin{figure}[!htb]
\centering
\includegraphics[width=0.9\textwidth]{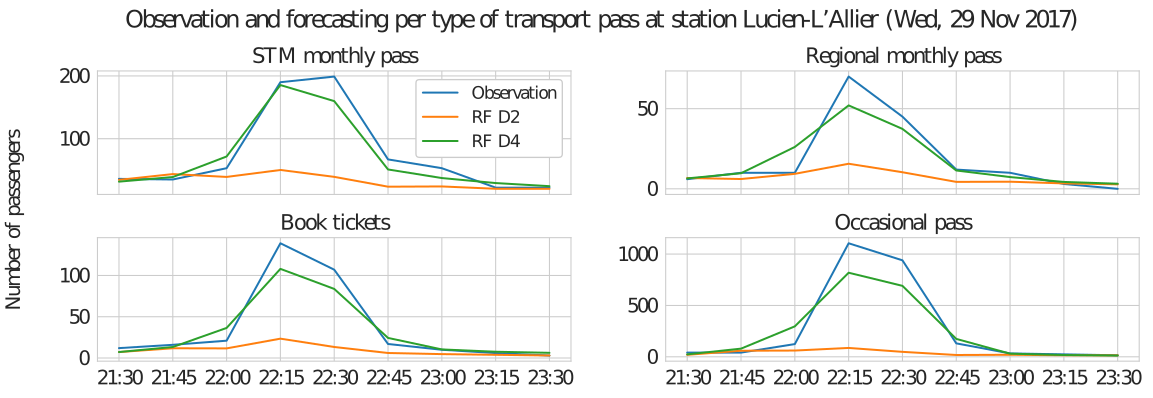}
\caption{Observation and forecasting of the passenger demand per type of ticket or pass at Lucien-L'Allier station, on Wednesday,  November 29, 2017. The information about the event is the following: start date and time, 2017-11-29, 19:30:00, end date and time Nan; station, Lucien-L'Allier;  and category, hockey.} \label{fig:obsvspred_perpass_lucienlallier}
\end{figure}

We can observe the impact of a hockey game on the passenger demand for each type of ticket or pass in Figure~\ref{fig:obsvspred_perpass_lucienlallier}. This event is described as beginning at 07:30 p.m.; however, the ending time is not defined. We can see that every random forest with the input data set D4 (day information, event and category information) is able to forecast with a good accuracy the increase in the number of passengers between 10:00 p.m. and 11:00 p.m. The type of ticket or pass used to travel that is the most impacted by the event is the occasional pass with an increase of 1000 passengers during the passenger demand peak at 10:15 p.m.

\section{Conclusion}
\label{sec:conclusion}

This paper has investigated the use of smart card, calendar and event data to forecast metro passenger demand per station with a long-term forecasting time horizon (until one year ahead) with fine-grained temporal resolution (15 minutes aggregation). We performed the forecasting task on real data (Montréal subway, Canada) by taking into account events in the city, such as concerts, hockey games, festivals, and so forth. The operational objectives were twofold: long-term forecasting can be useful for transport operators to adapt the transport supply and to adjust ticket availability to passenger demand. In this context, we have investigated the forecasting of the number of passengers per type of ticket or pass in addition to the forecasting of the global passenger demand.

We have proposed generic data shaping, allowing the use of contextual data (smart card, calendar and event data) as input for well-known regression models: basic, statistical and machine learning models. Global forecast analysis has proven that it is possible to obtain good long-term forecasting accuracy with fine-grained resolution even in the presence of events. The random forest model achieved the best forecasting results with the calendar information and event data as input. The forecasting results highlighted the importance of taking  event data into account during the forecasting of passenger demand, particularly during an event period. This study has also illustrated the value for transport operators to use one regression model per station to understand which features mostly impact the passenger demand per station. We have studied transport-related results to better understand which station is difficult to predict. In the same line of work, we have shown that, as expected, passenger demand depending on certain types of ticket or pass used to travel (book tickets and non-rechargeable smart cards) is more impacted by events and requires event data to be accurately forecasted.

We have also proposed a basic method to reproduce the impact of the global year-to-year trend on the forecasting results. These results have demonstrated the effectiveness of the trend method in addition to the data shaping and machine learning method for such forecasting tasks. Nevertheless, further work is required to investigate in detail the trend problem in the long-term prediction task. 
Future work could investigate a medium-term forecast that could be placed between a long-term forecast that requires only the use of data available well in advance and a short-term forecast that requires recent observations of passenger numbers (collection and analysis of near real-time data). For this purpose, the medium-term forecast model could take as inputs, in addition to long-term data (calendar and event information), medium-term features such as the trend of the number of passengers observed recently (e.g., in previous days, weeks or months).

If we take the case of a transport network with a constrained spatial grid that can be defined by stations (e.g., metro, bus, train), it will be possible to use the same forecast method as well as the same data formatting method. These forecasting methods are applicable provided that the same types of data (calendar and event data) are used. On the other hand, in the case of an unmeshed network as in~\cite{Zhang2017} (e.g., road traffic, free-floating bicycles, free-floating scooters), it will be necessary to spatially mesh the network in order to group the events into a number of fixed points of interest, as well as for the counting of transport flows, which will also have to be grouped into a fixed number of points.
Because it is desired to be applicable in other public transport systems of the world, the forecasting methodology presented in this study could definitely help to create high added value mobility services for citizens.

\section*{Acknowledgement}
This research was partially supported by Thales and the Natural Science and Engineering Research Council of Canada (NSERC). The authors wish to thank the transport organisation authority of Montreal (Société de transport de Montréal)
for providing the smart card and event data.  They also acknowledge the computer infrastructure support from the IVADO and the Quebec Research Funds (FRQNT, FRQSC).


\end{document}